\documentclass[pre,aps,reprint,longbibliography]{revtex4-1}

\usepackage{natbib}

\usepackage{amsmath}
\usepackage{amssymb}
\usepackage{mathtools}
\usepackage{graphicx}
\usepackage{subfigure}
\usepackage{algorithm}
\usepackage{algpseudocode}
\usepackage{bm}
\usepackage{color}
\usepackage{xcolor}

\usepackage{hyperref}

\begin{document}

\title{Cycle-tree guided attack of random $K$-core: Spin glass model and efficient message-passing algorithm}

\author{Hai-Jun Zhou$^{1,2,3}$}
\email{zhouhj@itp.ac.cn}

\affiliation{
  $^1$CAS Key Laboratory for Theoretical Physics, Institute of Theoretical Physics, Chinese Academy of Sciences, Beijing 100190, China \\
  $^2$MinJiang Innovative Center for Theoretical Physics, MinJiang University, Fuzhou 350108, China \\
  $^3$School of Physical Sciences, University of Chinese Academy of Sciences, Beijing 100049, China
}

\date{\today}

\begin{abstract}
  The $K$-core of a graph is the maximal subgraph within which each vertex is connected to at least $K$ other vertices. It is a fundamental network concept for understanding threshold cascading processes with a discontinuous percolation transition. A minimum attack set contains the smallest number of vertices whose removal induces complete collapse of the $K$-core.  Here we tackle this prototypical optimal initial-condition problem from the spin-glass perspective of cycle-tree maximum packing and propose a cycle-tree guided attack ({\tt CTGA}) message-passing algorithm. The good performance and time efficiency of {\tt CTGA} are verified on the regular random and Erd\"os-R\'enyi random graph ensembles. Our central idea of transforming a long-range correlated dynamical process to static structural patterns may also be instructive to other hard optimization and control problems.
  \\
  Key words: K-core collapse, spin glass model, tree packing, optimal initial condition, random graph
  \\
  PACS: 05.70.Fh (Phase transitions; general studies); 89.90.+n (Other topics in areas of applied and interdisciplinary physics); 02.10.Ox (Combinatorics; graph theory); 75.10.Nr (Spin-glass and other rnadom models)
\end{abstract}

\maketitle

\section{Introduction}

The $K$-core of a graph $G$ is the maximal subgraph within which every vertex is linked to $K$ or more other vertices. This concept originally was introduced to understand collective behaviors in ferromagnets~\cite{Pollak-Riess-1975} and social systems~\cite{Granovetter-1978,Seidman-1983}, but later found applications in diverse problems such as structural glass~\cite{Schwarz-Liu-Chayes-2006,Sellitto-2013}, ecological system~\cite{Morone-etal-2018} and influential spreader identification~\cite{Kitsak-etal-2010}. If non-empty the $K$-core of a graph is unique, but there are many ways to destroy it. Following the deletion of an initial set of vertices, some other vertices may leave spontaneously if they have fewer than $K$ neighbors in the $K$-core, which may in turn cause more vertices to leave. The $K$-core percolation transition caused by random deletion of vertices is discontinuous in general for $K \geq 3$~\cite{Chalupa-Leath-Reich-1979,Pittel-etal-1996,Dorogovtsev-etal-2006,Baxter-etal-2010,Shrestha-Moore-2014,Rizzo-2018}. A much more challenging issue is the worst-case robustness, that is, the minimum-sized attack that is sufficient to induce complete collapse of the $K$-core~\cite{Altarelli-Braunstein-DallAsta-Zecchina-2013,Altarelli-Braunstein-DallAsta-Zecchina-2013b,Zhou-2013,Guggiola-Semerjian-2015,Yuan-etal-2016,Pei-etal-2017,Schmidt-etal-2018,Wang-etal-2020}. This optimization task is a problem of optimal initial condition.

During the last two decades, the mean field spin glass theory has achieved huge success in tackling random constraint satisfaction and combinatorial optimization problems~\cite{Mezard-etal-2002,Krzakala-etal-PNAS-2007,Mezard-Montanari-2009,Zhou-2015}. When applying this theoretical framework to irreversible threshold dynamics such as $K$-core disruption, the standard approach has been to map the cascade transmission into a Potts model with $T\! +\! 1$ states $s_i \in \{0, 1, \ldots, T\}$, where $s_i\! =\! 0$ means that vertex $i$ is deleted externally (an initial driver), and $s_i \! =\!  t$ $(>\! 0)$ means that vertex $i$ is affected at the $t$-th time step following the threshold dynamics~\cite{Altarelli-Braunstein-DallAsta-Zecchina-2013,Altarelli-Braunstein-DallAsta-Zecchina-2013b,Guggiola-Semerjian-2015,Braunstein-etal-2016,Zhou-2016b,Zhao-Zhou-2016}. Although this mapping is straightforward, the resulting model is computationally heavy when the state dimension $T$ becomes large, and a moderate cut-off of $T \! \approx \! 10^2$ has to be imposed to make numerical computation tolerable. On the other hand, an optimal initial condition usually implies a long time $T$ to finish the threshold dynamics, even up to $T\! \propto \! O(N)$ if the global graph topology is a torus. 

For the special case of threshold $K\! = \! 2$, the minimum attack problem is identical to the minimum feedback vertex set problem of destroying all the loops in the graph. After the initial driven vertices are all deleted from the graph, the residual graph becomes a collection of mutually disconnected tree components. This special structural feature could be exploited to tackle the $2$-core problem as the densest packing problem of tree components~\cite{Zhou-2013}, and it also leads to a very efficient message-passing algorithm with only three coarse-grained vertex states for $2$-core attack and graph dismantling~\cite{Mugisha-Zhou-2016,Li-etal-2021}. But for general $K \geq 3$, the graph after deleting all the initial driven vertices is not a forest but is still rich in loops, rendering  this simple tree-packing model invalid for general $K$-core.

In the present work, we demonstrate that by allowing the different tree components to be adjacent to each other and allowing for additional edges within each tree component, actually the tree-packing model~\cite{Zhou-2013} can be extended to the more difficult $K$-core attack problem. We describe in detail the simplest implementation of this extension and the inspired message-passing algorithm. Our algorithm is very efficient as it only needs four coarse-grained vertex states for any arbitrarily large graph, and it outperforms the state-of-the-art greedy algorithm~\cite{Schmidt-etal-2018} on regular random and Erd\"os-R\'enyi random graphs. A easily conceivable refinement of our extended packing model is also briefly mentioned at the end of this paper and will be thoroughly investigated in a follow-up publication.  Our central idea of transforming a long-range correlated dynamical process to special static structural patterns may offer a new perspective to other related hard optimization and control problems~\cite{Liu-Barabasi-2016}. $K$-core percolation is a popular topic in network science, and our message-passing algorithm may also be helpful for identifying influential nodes in complex networks.

\section{Minimum attack set}

Consider a simple graph $G$ of $N$ vertices and $M$ undirected edges. The vertices are indexed by positive integers $i, j, k, \ldots$, and two vertices $i$ and $j$ are neighbors if there is an edge $(i, j)$ between them. The neighborhood of vertex $i$ is the set $\partial i \equiv \{j: (i, j) \in G\}$ of all its neighbors, and the cardinality of this set is its degree $d_i$. Without loss of generality, we assume initially the whole graph is a $K$-core, $d_i \geq K$ for all the vertices $i$. The vertices are strongly mutually dependent in the $K$-core. The deletion of a single vertex may sometimes cause an extensive damage to the $K$-core through the cascade of local failures, with vertices leaving the $K$-core when they have fewer than $K$ insider neighbors~\cite{Wang-etal-2020}. Leveraging on this dependence effect, the optimization task on graph $G$ is to construct a vertex set of minimum cardinality (a minimum attack set $\Gamma$), such that if all the vertices in $\Gamma$ are deleted, the residual graph of $G$ will have no $K$-core (in other words, the original $K$-core will completely collapse).

The $K$-core minimum attack problem is intrinsically hard. Many heuristic algorithms have been designed to solve this problem approximately. One simple and effective recipe is CoreHD~\cite{Zdeborova-etal-2016}, which iteratively deletes one of the most connected (highest-degree) vertices in the remaining $K$-core. The performance of CoreHD could be further improved by deleting at each decimation step a vertex which is itself highly connected but whose neighbors have relatively low degrees, namely a vertex $i$ with the maximum value of $d_i\! -\! d_i^{(n)}$ with $d_i^{(n)}$ being the average degree of the neighbors~\cite{Schmidt-etal-2018}. Extensive simulation results have demonstrated the superior performance of this Weak-Neighbors ({\tt WN}) algorithm over the more sophisticated collective-information message-passing algorithm {\tt CI-TM}~\cite{Pei-etal-2017}. (Empirical evidence of \cite{Pei-etal-2017} indicated that {\tt CI-TM} performs better than several centrality-based algorithms and that it is comparable in performance with a message-passing algorithm {\tt Min-Sum} inspired by the conventional spin glass model~\cite{Altarelli-Braunstein-DallAsta-Zecchina-2013,Altarelli-Braunstein-DallAsta-Zecchina-2013b}.) Very remarkably, the attack sets $\Gamma$ constructed by {\tt WN} for random regular graphs are quite close to the known rigorous lower-bound values~\cite{Bau-Wormald-Zhou-2002} or to the non-rigorous minimum values conjectured by the mean field theory~\cite{Guggiola-Semerjian-2015}. We now build a new heuristic algorithm, Cycle-Tree-Guided-Attack ({\tt CTGA}), which significantly outperforms {\tt WN} on the tested random graphs. Our algorithm takes the global structural property of the graph into account by message passing, and its time complexity could be made roughly linear in $N$. 
 
\section{Cycle-tree packing model}

\begin{figure}
  \centering
   \includegraphics[width=1.0\linewidth]{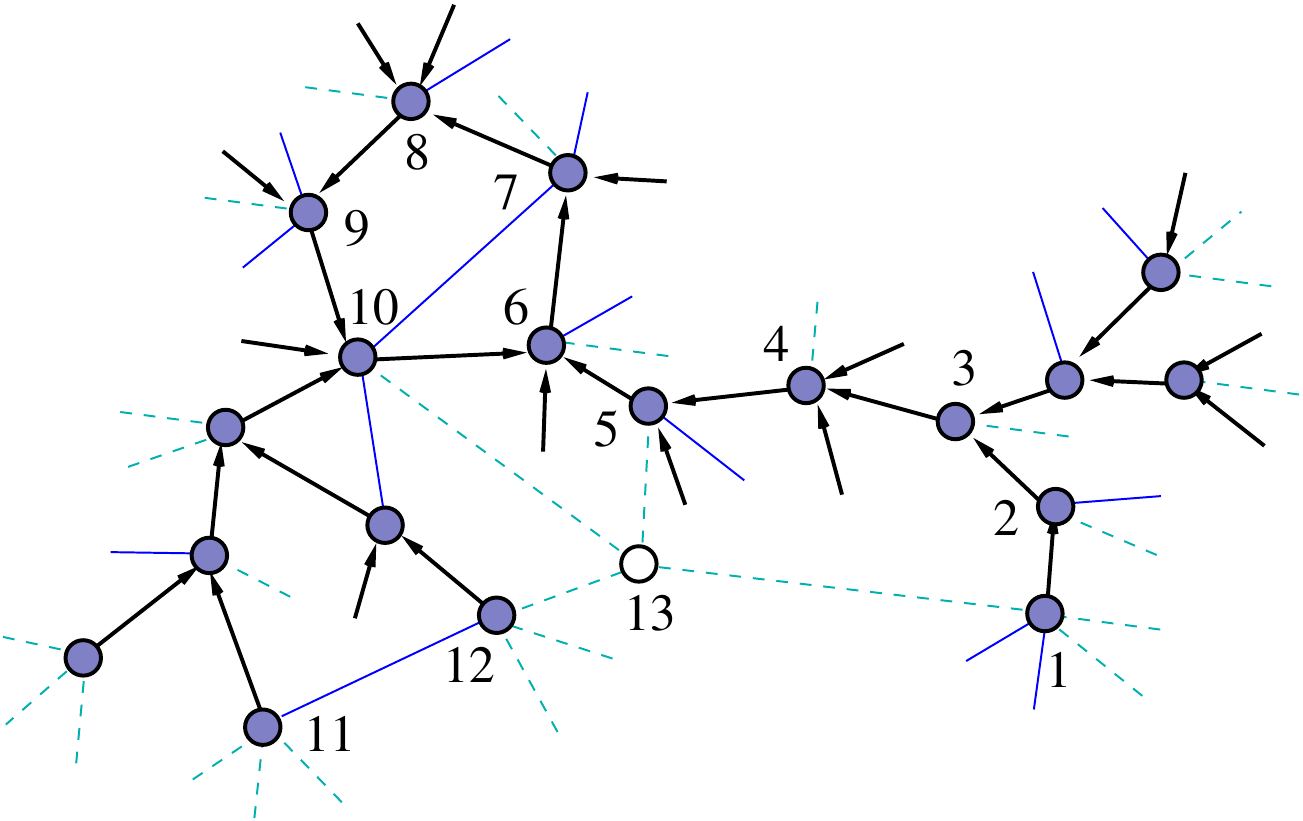}
   \caption{
     \label{fig:example}
     Part of a directed cycle-tree, showing a directed cycle of five vertices ($6\! \rightarrow \! 7\! \rightarrow\! 8\! \rightarrow \! 9\! \rightarrow \! 10 \! \rightarrow \! 6$) and two directed tree branches attached to it.  Filled circles indicate occupied vertices and open circles empty ones. Edges between occupied vertices are drawn as solid lines, those attached to an empty vertex such as $13$ are drawn as dashed lines.  An arrow-decorated solid edge, say $5\! \rightarrow\! 6$, indicates the state of vertex $5$ being $A_5\! =\! 6$. An occupied vertex has one arrow-decorated edge pointing outward and it allows at most $K\! - \! 2$ ordinary arrow-free solid edges (here $K\! = \! 4$). 
  }
\end{figure}

We assign a discrete state $A_i \in \{0, i\} \cup \partial i$ of $d_i\! +\! 2$ values to each vertex $i$ of graph $G$. Vertex $i$ is said to be empty if $A_i\! =\! 0$, otherwise it is said to be occupied. If an occupied vertex $i$ has state $A_i \! = \! i$, we call it a root, otherwise we represent the state $A_i\! =\! j$ ($\in\! \partial i$) by an arrow on edge $(i, j)$ pointing to $j$, giving this edge a direction (Fig.~\ref{fig:example}). Notice that every occupied non-root vertex $i$ has exactly one out-going edge, with the other edges either arrow-free or pointing to it. Let us draw the edges between occupied vertices as solid lines and the edges attached to the empty vertices as dashed lines (Fig.~\ref{fig:example}). Each vertex $i$ imposes the following constraint to the states of itself and its neighbors: (1) if $i$ is empty, none of its neighbors shall point to it; (2) if $i$ points to $j$ ($A_i\! =\! j$), then vertex $j$ shall be occupied but not be pointing to $i$ ($A_j \neq 0, i$) and the total number of other occupied neighbors not pointing to $i$ shall be at most $K\!-\! 2$; (3) if $i$ is a root, all its occupied neighbors shall point to it. In mathematical terms this means the condition $C_i = 1$, with the constraint factor $C_i \in \{0, 1\}$ being
\begin{eqnarray}
  & & C_i \bigl( A_i, \underline{A}_{\partial i}\bigr) =
  \delta_{A_i}^{0} \prod\limits_{j\in \partial i} \bigl( \delta_{A_j}^0 +
  \delta_{A_j}^j + \Delta_{A_j}^{\backslash i} \bigr) +
  \nonumber \\
  & & \quad 
  \sum\limits_{j\in \partial i} \delta_{A_i}^{j} \bigl( \delta_{A_j}^j +
  \Delta_{A_j}^{\backslash i} \bigr)  \prod\limits_{k\in \partial i\backslash j}
  \bigl( \delta_{A_k}^0 + \delta_{A_k}^i + \Delta_{A_k}^{\backslash i} \bigr) \times
  \nonumber \\
  & & \quad  
  \Theta \bigl( K - 2 - \sum\limits_{l\in \partial i\backslash j}
  \Delta_{A_l}^{\backslash i} \bigr) + \delta_{A_i}^{i}
  \prod\limits_{j\in \partial i} \bigl( \delta_{A_j}^0 + \delta_{A_j}^i \bigr)
  \; ,
  \label{eq:Vfactor}
\end{eqnarray}
where $\underline{A}_{\partial i} \! \equiv\! \{A_j: j \in \partial i\}$ contains the states of all the neighboring vertices, $\delta_{m}^{n} =1$ if $m\! =\! n$ and $=0$ otherwise, the set $\partial i\backslash j$ contains all the neighbors of vertex $i$ except for $j$,  $\Delta_{A_j}^{\backslash i} \equiv \sum_{m \in \partial j\backslash i} \delta_{A_j}^m \in \{0, 1\}$ indicates whether vertex $j$ points to one of its neighbors other than $i$, and the step function $\Theta( x ) = 1$ if $x \! \geq\! 0$ and $ = 0$ if $x\! < \! 0$. Under all these vertex constraints, the occupied vertices and the arrow-decorated edges form a set of directed trees and cycle-trees~\cite{Zhou-2013}. A directed tree is a connected subgraph of $n$ vertices and $n\! - \! 1$ arrow-decorated edges, and a cycle-tree has $n\! \geq \! 3$ vertices and $n$ arrow-decorated edges (one directed cycle plus a set of directed tree branches pointing to this cycle, Fig.~\ref{fig:example}).

There could be arrow-free solid edges between vertices of the same (cycle-)tree or between two different (cycle-)trees. But the number of such arrow-free solid edges attached to each occupied vertex is at most $K\! -\! 2$, and consequently this vertex will not be able to remain in the $K$-core if all its pointing-in arrow-decorated edges are deleted. If all the empty vertices are deleted from graph $G$, the remaining occupied vertices will not sustain a $K$-core but instead is guaranteed to collapse, starting from the leaf vertices and propagating to the roots or central cycles (in the rare cases of a cycle being marginally stable, one of its vertices needs to be manually deleted). The problem of packing the vertices into trees and cycle-trees to cover graph $G$ is therefore closely related to the $K$-core attack problem, with the empty vertices form a $K$-core attack set (but very likely it is \emph{not} minimum in size).

To encourage more occupied vertices in this packing problem, we impose a penalty factor $e^{-\beta}$ with an inverse temperature $\beta$ to each empty vertex. The partition function of the spin glass model is then
\begin{equation}
  Z(\beta) = \sum\limits_{A_1, \ldots, A_N} \prod\limits_{i=1}^{N}
  \Bigl[  e^{-\beta \delta_{A_i}^0}
  C_i\bigl( A_i, \underline{A}_{\partial i} \bigr) \Bigr] \; .
\end{equation}
Based on this partition function we could evaluate the marginal probability $q_i^0$ of vertex $i$ being empty. We employ $q_i^0$ as an influence index to iteratively delete vertices from the remaining $K$-core (these marginal empty probabilities are updated after each perturbation of vertex deletions).

\section{Cycle-tree guided attack}

We adopt the now standard cavity method of statistical physics~\cite{Mezard-Montanari-2009,Zhou-2015} to efficiently estimate $q_i^0$. For every edge $(i, j)$ we denote by $q_{i\rightarrow j}^{A_i, A_j}$ the marginal probability of the states $A_i$ and $A_j$ in the cavity system which lifts the constraint $C_j$ from vertex $j$. A self-consistent belief-propagation (BP) equation can be easily written down for this cavity probability (e.g., by following the same procedure of \cite{Zhou-2016b}). For computing $q_i^0$ we actually need only to update four cavity probabilities $Q_{i\rightarrow j}^{c_i}$ for four effective states $c_i \in \{0,1,2,3\}$, and the marginal empty probability $q_i^0$ of vertex $i$ is then estimated using these merged cavity probabilities as
\begin{eqnarray}
  & & q_i^0  =
  e^{-\beta} \prod\limits_{j\in \partial i} \bigl(Q_{j\rightarrow i}^0 +
  Q_{j\rightarrow i}^2 \bigr)  \Big/ \Bigl\{
  e^{-\beta} \prod\limits_{j\in \partial i} \bigl(Q_{j\rightarrow i}^0 +
  Q_{j\rightarrow i}^2 \bigr) 
  \nonumber \\
 & & \quad  +  \sum\limits_{j\in \partial i}
  Q_{j\rightarrow i}^{2}
  \prod\limits_{k\in \partial i\backslash j}
  \sum\limits_{c_k}
  \bigl(\delta_{c_k}^0 Q_{k\rightarrow i}^0
  + \delta_{c_k}^1 Q_{k\rightarrow i}^1
  +\delta_{c_k}^3 Q_{k\rightarrow i}^3 \bigr)
  \nonumber \\
  & & \quad  \times 
  \Theta\bigl(K-2 -
  \sum{_{l\in \partial i\backslash j}} \delta_{c_l}^3 \bigr)
  + \prod\limits_{j\in \partial i} \bigl(Q_{j\rightarrow i}^0 +  
   Q_{j\rightarrow i}^1 \bigr)   \Bigr\}
   \; .
  \label{eq:q0}
\end{eqnarray}

The definitions of these four merged probabilities $Q_{i\rightarrow j}^{c_i}$ (e.g., $Q_{i\rightarrow j}^0$ corresponds to vertex $i$ being empty and  $Q_{i\rightarrow j}^1$ corresponds to vertex $i$ pointing to vertex $j$) and the corresponding coarse-grained BP equation are
\begin{subequations}
  \label{eq:BP}
  \begin{align}
     & Q_{i\rightarrow j}^{0}  \equiv \frac{1}{d_j+1}
    \bigl( q_{i\rightarrow j}^{0,0} + q_{i\rightarrow j}^{0,j} +
    \sum{_{l \in \partial j\backslash i}} q_{i\rightarrow j}^{0, l} \bigr)
    \nonumber \\
    & \quad = \frac{1}{z_{i\rightarrow j}} e^{-\beta}
    \prod\limits_{k\in \partial i\backslash j}
    \bigl( Q_{k\rightarrow i}^{0} + Q_{k\rightarrow i}^{2} \bigr) \; ,
    \\
    & Q_{i\rightarrow j}^{1}  \equiv \frac{1}{d_j} \bigl( q_{i\rightarrow j}^{j,j} +
    \sum{_{l \in \partial j\backslash i}} q_{i\rightarrow j}^{j, l} \bigr)
    \nonumber \\
    & \quad = \frac{1}{z_{i\rightarrow j}}
    \prod\limits_{k\in \partial i\backslash j} \sum\limits_{c_k}
    \bigl(\delta_{c_k}^0 Q_{k\rightarrow i}^{0} + \delta_{c_k}^1
    Q_{k\rightarrow i}^{1} + \delta_{c_k}^3  Q_{k\rightarrow i}^{3} \bigr)
    \nonumber \\
    & \quad\quad \quad \quad \times
        \Theta\bigl( K - 2 -
    \sum{_{k\in \partial i\backslash j}} \delta_{c_k}^3 \bigr)
    \; ,
    \\
   & Q_{i\rightarrow j}^{2} \equiv  \frac{1}{2} \bigl( q_{i\rightarrow j}^{i,0} +
    q_{i\rightarrow j}^{i, i}+\sum{_{k\in \partial i\backslash j}}
    ( q_{i\rightarrow j}^{k, 0} + q_{i\rightarrow j}^{k, i} ) \bigr)
    \nonumber \\
    & \quad = \frac{1}{z_{i\rightarrow j}} \Bigl\{
    \prod\limits_{k\in \partial i\backslash j} \bigl( Q_{k\rightarrow i}^0
    + Q_{k\rightarrow i}^1 \bigr)    + \sum\limits_{k\in \partial i\backslash j}
    Q_{k\rightarrow i}^{2} \times 
   \nonumber \\
    & \quad \quad  
    \prod\limits_{l\in \partial i\backslash j, k} \sum\limits_{c_l}
    \bigl(\delta_{c_l}^0 Q_{l\rightarrow i}^{0} + \delta_{c_l}^1
    Q_{l\rightarrow i}^{1} + \delta_{c_l}^3  Q_{l\rightarrow i}^{3} \bigr)
    \nonumber \\
    &   \quad \quad \quad \times
      \Theta\bigl(K-2 -
    \sum{_{l\in \partial i\backslash j, k}} \delta_{c_l}^3 \bigr)
    \Bigr\}
    \; ,
    \\
    & Q_{i\rightarrow j}^{3}  \equiv \frac{1}{d_j - 1}
    \sum{_{k\in \partial i\backslash j}} \sum{_{l\in \partial j\backslash i}}
    q_{i\rightarrow j}^{k, l}
    \nonumber \\
    & \quad  = \frac{1}{z_{i\rightarrow j}}
    \sum\limits_{k\in \partial i\backslash j}
    Q_{k\rightarrow i}^{2} 
    \prod\limits_{l\in \partial i\backslash j, k}
    \sum\limits_{c_l}
    \bigl( \delta_{c_l}^0 Q_{l\rightarrow i}^{0}
    + \delta_{c_l}^1  Q_{l\rightarrow i}^{1}
    \nonumber \\
    & \quad \quad \quad
    + \delta_{c_l}^3  Q_{l\rightarrow i}^{3} \bigr)
    \Theta\bigl(K-3 -
     \sum{_{l\in \partial i\backslash j, k}} \delta_{c_l}^3 \bigr) 
     \; ,
  \end{align}
\end{subequations}
where $\partial i\backslash j, k$ is the residual of set $\partial i$ after removing $j$ and $k$, and $z_{i\rightarrow j}$ is the normalization constant ensuring $(d_j+1) Q_{i\rightarrow j}^{0} + d_j Q_{i\rightarrow j}^{1} + 2 Q_{i\rightarrow j}^{2} + (d_j-1) Q_{i\rightarrow j}^3 = 1$.

\begin{figure*}
  \centering
  \subfigure[]{
    \label{fig:CTGAD7K3}
    \includegraphics[angle=270,width=0.469\linewidth]{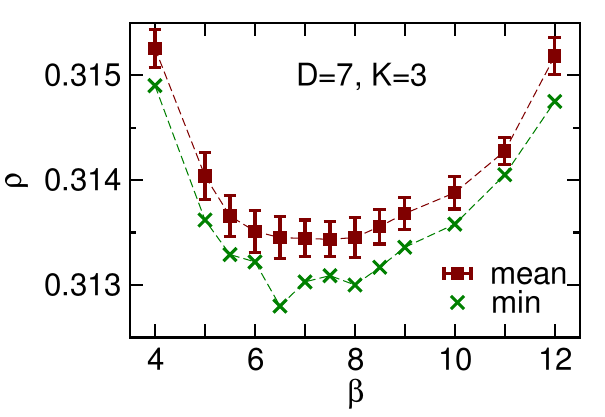}
  }
  \subfigure[]{
    \label{fig:CTGAD10K4}
    \includegraphics[angle=270,width=0.469\linewidth]{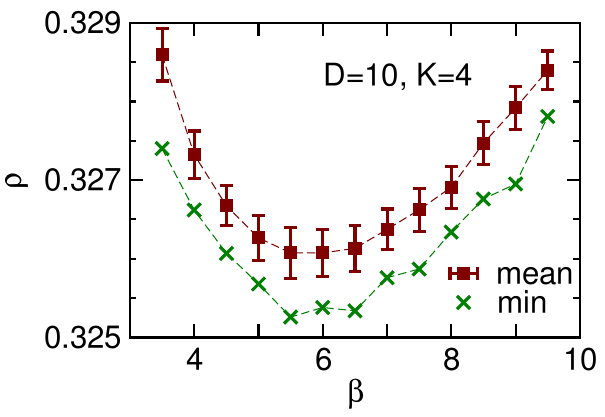}
  }
  \caption{
    \label{fig:CTGA}
    Performance of {\tt CTGA} on a single large RR graph instance ($N=10^5$). The minimum, mean and standard deviation of the relative attack sizes $\rho$ at each inverse temperature $\beta$ are estimated over $50$ independent runs.  (a) $D=7$, $K=3$; (b) $D=10$, $K=4$. In comparison, the mean value of $\rho$ obtained by {\tt WN} is $0.3206$ for (a) and $0.3331$ for (b).
  }
\end{figure*}

We implement the {\tt CTGA} algorithm in the following simple way. At each decimation step we first run the BP iteration (\ref{eq:BP}) on the remaining $K$-core for a small number of repeats and estimate the empty probability for each of its vertices $i$ using (\ref{eq:q0}), then we delete the vertex (say $j$) with the maximum value of $q_j^0$ from the $K$-core and add it to the attack set $\Gamma$, simplifying the graph after each vertex deletion. When the $K$-core completely disappears, the final relative size $\rho$ (with respect to $N$) of the attack set $\Gamma$ is then reached.  When $K \! = \! 2$, {\tt CTGA} is the same as the algorithms developed in \cite{Zhou-2013,Li-etal-2021} for tackling the minimum feedback vertex set problem.

\begin{figure*}
  \centering
  \subfigure[]{
    \label{fig:ComparisonRho}
    \includegraphics[angle=270,width=0.469\linewidth]{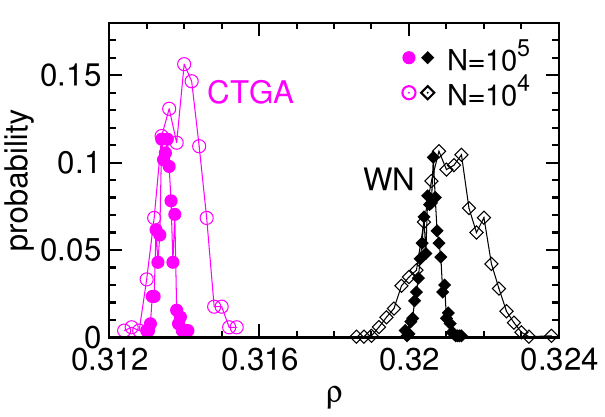}
  }
  \subfigure[]{
    \label{fig:ComparisonEvol}
    \includegraphics[angle=270,width=0.469\linewidth]{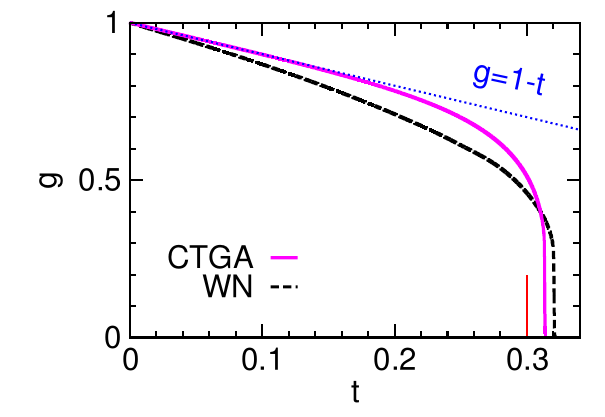}
  }
  \subfigure[]{
    \label{fig:ComparisonTime}
    \includegraphics[angle=270,width=0.469\linewidth]{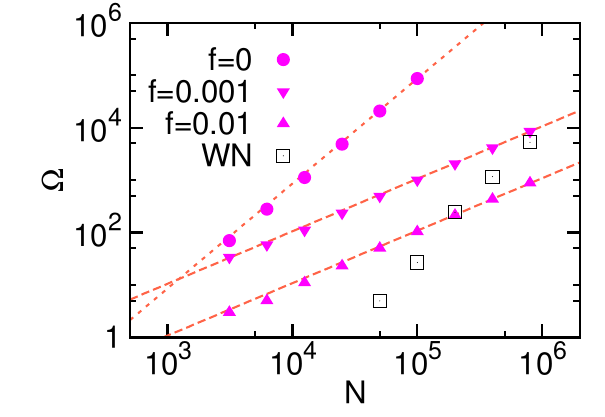}
  }
  \subfigure[]{
    \label{fig:ComparisonSize}
    \includegraphics[angle=270,width=0.469\linewidth]{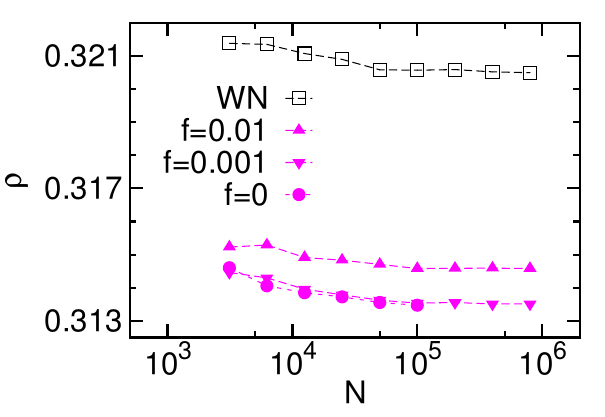}
  }
  \caption{
    \label{fig:Comparison}
    Comparing {\tt CTGA} with $\beta\! = \! 7.5$ and {\tt WN} on RR graphs of different sizes $N$ ($D\! = \! 7$, $K\! =\! 3$). (a) The probability of relative attack size $\rho$ for a single graph instance ($N \! =\! 10^4$ or $10^5$). (b) Evolution of the relative $K$-core size $g$ with the relative number $t$ of attacked vertices ($N\!= \! 10^5$; vertical line marks the conjectured minimum $\rho$~\cite{Guggiola-Semerjian-2015}, dotted line is $g\! =\! 1\! -\! t$). (c) and (d): Total time $\Omega$ (in seconds) {\tt CTGA} and {\tt WN} spend on the decimation process (c), and the final relative attack size $\rho$ (d). Each point in (c) and (d) is obtained on a single graph instance with $1000$ independent runs; $f$ is the fraction of $K$-core vertices deleted in each {\tt CTGA} decimation step ($f\! = \! 0$ indicates deleting only a single vertex); dashed lines in (c) mark the linear scaling $\Omega \! \propto\! N$, dotted one marks the quadratic scaling $\Omega\! \propto\! N^2$. 
  }
\end{figure*}

\section{Numerical experiments}

We test the performance of {\tt CTGA} on regular random (RR) graphs. Each vertex in such a graph is connected to the same number $D$ of randomly chosen neighbors. These homogeneous and structureless graphs are most challenging, as there are no local structural clues to guide the $K$-core attack process. {\tt CTGA} significantly outperforms {\tt WN} on this graph ensemble. For example, when $D\! =\! 7$ and $K\! =\! 3$ {\tt CTGA} achieves complete $K$-core collapse after attacking a fraction of vertices as few as $\rho\! \approx\! 0.3130$ while {\tt WN} only succeeds at $\rho \! \geq \! 0.3200$; when $D\! =\! 10$ and $K\! =\! 4$, {\tt CTGA} succeeds at minimal value $\rho\! \approx\! 0.3255$  (Fig.~\ref{fig:CTGA}) while {\tt WN} needs at least $\rho\! \approx\! 0.3321$. The performance of {\tt CTGA} has a slight dependence on the inverse temperature $\beta$ and it works best at intermediate $\beta$ values (if $\beta$ is too large the system is deep in the spin glass phase and the BP iterations are strongly divergent).  If $\beta$ is fixed during the whole decimation process, the optimal value is $\beta \! \approx\! 7.5$ for $D\! =\! 7$ and $K\! =\! 3$ and it is $\beta \! \approx\! 6.0$ for $D\! =\! 10$ and $K\! =\! 4$.

To compare the scaling behaviors of {\tt CTGA} and {\tt WN}, we consider a set of RR graphs with different sizes $N$ and the same degree $D$ (Fig.~\ref{fig:Comparison}). For each graph instance we repeat both the {\tt CTGA} and {\tt WN} processes at least $1000$ times to obtain two histograms of the relative attack size $\rho$, which are always well separated and are insensitive to $N$ (Fig.~\ref{fig:ComparisonRho}). Examining the $K$-core relative size $g$ as a function of the fraction $t$ of already attacked vertices, we find that {\tt CTGA} does not show any cooperative effect during an extended initial stage ($g = 1-t$) while {\tt WN} induces an efficient initial reduction of the $K$-core size; this ``defeat at the start point'' is a trade-off made by {\tt CTGA} for building potential for severe cascading damages at the later evolution stage to ``win at the end point'' (Fig.~\ref{fig:ComparisonEvol}). We have tried to combine the excellent initial efficiency of {\tt WN} with the final acceleration of {\tt CTGA}, but such a {\tt WN}-then-{\tt CTGA} hybrid algorithm turns out to be worse than pure {\tt CTGA} in performance.

Similar to {\tt WN}, the {\tt CTGA} algorithm has an approximately $O(N^2)$ time complexity if only a single vertex is selected and deleted from the $K$-core at each decimation step. We can select and delete a small fraction $f$ of the $K$-core vertices to reduce the time complexity to approximately $O(N)$ (Fig.~\ref{fig:ComparisonTime}). This acceleration does not compromise the algorithmic performance as long as $f$ is small (e.g., $f=0.001$, Fig.~\ref{fig:ComparisonSize}), making {\tt CTGA} applicable for very large graph instances. 

We have also applied {\tt CTGA} on the more heterogeneous Erd\"os-R\'enyi graphs whose $M$ edges are randomly picked from all the $N(N-1)/2$ candidate edges. As expected, {\tt CTGA} performs distinctly better than {\tt WN} on this second random graph ensemble.

\section{Outlook}

The cycle-tree packing model offered a new solution concept and an efficient {\tt CTGA} heuristic for the  $K$-core minimum attack problem. The central idea was to represent an irreversible threshold dynamics by certain static patterns (here cycle-trees) in the graph. This dynamic-static transformation (or conversion) may also be fruitful for some even harder optimization problems, such as general minimum attack problems of directed graphs~\cite{Zhao-2017} and the feedback vertex and arc set problems of directed graphs~\cite{Zhou-2016b,Xu-Zhou-2017}.

There is still much room for improving the {\tt CTGA} algorithm. As demonstrated in Fig.~\ref{fig:ComparisonEvol}, the best result $\rho \! \approx \! 0.3130$ achieved by {\tt CTGA} at $K\! = \! 3$ on large RR graphs of degree $D \! = \! 7$ is noticeably larger than the theoretical minimum $\rho \! \approx \! 0.3001$ (which may only be a lower-bound) conjectured by the non-rigorous mean field theory~\cite{Guggiola-Semerjian-2015}. Informed by the $\beta$-dependence shown in Fig.~\ref{fig:CTGA}, maybe it will be rewarding to train an adaptable $\beta$ for each individual decimation step through reinforcement learning techniques. On the more fundamental level, the present cycle-tree packing constraint (\ref{eq:Vfactor}) is a bit too stringent, as it prohibits every occupied vertex from having more than $K\! -\! 2$ arrow-free solid edges. The predicted ground-state energy density of this model for the RR graph ensemble of $D=7$ at $K=3$ is quite high, $\rho\approx 0.376$. Maybe we should distinguish between two (or even more) types of occupied vertices: an $e$-type (earlier-unstable) vertex allows at most $K\! - \! 2$ attached arrow-free solid edges as in the present work, while an $l$-type (later-unstable) vertex tolerates more arrow-free solid edges as long as no more than $K\! -\! 2$ of them are linking to $l$-type peers. Our preliminary theoretical results indicate that the ground-state energy density of the RR graph ensemble of $D=7$ at $K=3$ will be decreased to $\rho \approx 0.332$ by considering two types of occupied vertices. A thorough investigation of this refined but more complicated model with nice coarse-grained effective states and its algorithmic implications will be the subject of a separate technical publication. It will also be interesting to adapt the technique of Ref.~\cite{Altarelli-Braunstein-DallAsta-Zecchina-2013} to consider the $\beta \rightarrow \infty$ limit of the cycle-tree algorithms and compare their performances with that of the {\tt Min-Sum} algorithm. Application of the {\tt CTGA} algorithm to real-world complex network instances may motivate further adaptations to different network ensembles~\cite{Li-etal-2021}.

\begin{acknowledgments}
  The author thanks Dr.~Jin-Hua Zhao for helpful discussions. This study was partly supported by the National Natural Science Foundation of China Grants No.~11975295 and No.~12047503, and the Chinese Academy of Sciences Grants No.~QYZDJ-SSW-SYS018 and No.~XDPD15. Numerical simulations were carried out at the Tianwen clusters of ITP-CAS and the BSCC-A3 cluster of Beijing Super Cloud Computing Center (PARATERA Tech., https://www.paratera.com/).
\end{acknowledgments}


\begin{thebibliography}{35}%
\makeatletter
\providecommand \@ifxundefined [1]{%
 \@ifx{#1\undefined}
}%
\providecommand \@ifnum [1]{%
 \ifnum #1\expandafter \@firstoftwo
 \else \expandafter \@secondoftwo
 \fi
}%
\providecommand \@ifx [1]{%
 \ifx #1\expandafter \@firstoftwo
 \else \expandafter \@secondoftwo
 \fi
}%
\providecommand \natexlab [1]{#1}%
\providecommand \enquote  [1]{``#1''}%
\providecommand \bibnamefont  [1]{#1}%
\providecommand \bibfnamefont [1]{#1}%
\providecommand \citenamefont [1]{#1}%
\providecommand \href@noop [0]{\@secondoftwo}%
\providecommand \href [0]{\begingroup \@sanitize@url \@href}%
\providecommand \@href[1]{\@@startlink{#1}\@@href}%
\providecommand \@@href[1]{\endgroup#1\@@endlink}%
\providecommand \@sanitize@url [0]{\catcode `\\12\catcode `\$12\catcode
  `\&12\catcode `\#12\catcode `\^12\catcode `\_12\catcode `\%12\relax}%
\providecommand \@@startlink[1]{}%
\providecommand \@@endlink[0]{}%
\providecommand \url  [0]{\begingroup\@sanitize@url \@url }%
\providecommand \@url [1]{\endgroup\@href {#1}{\urlprefix }}%
\providecommand \urlprefix  [0]{URL }%
\providecommand \Eprint [0]{\href }%
\providecommand \doibase [0]{http://dx.doi.org/}%
\providecommand \selectlanguage [0]{\@gobble}%
\providecommand \bibinfo  [0]{\@secondoftwo}%
\providecommand \bibfield  [0]{\@secondoftwo}%
\providecommand \translation [1]{[#1]}%
\providecommand \BibitemOpen [0]{}%
\providecommand \bibitemStop [0]{}%
\providecommand \bibitemNoStop [0]{.\EOS\space}%
\providecommand \EOS [0]{\spacefactor3000\relax}%
\providecommand \BibitemShut  [1]{\csname bibitem#1\endcsname}%
\let\auto@bib@innerbib\@empty
\bibitem [{\citenamefont {Pollak}\ and\ \citenamefont
  {Riess}(1975)}]{Pollak-Riess-1975}%
  \BibitemOpen
  \bibfield  {author} {\bibinfo {author} {\bibfnamefont {M.}~\bibnamefont
  {Pollak}}\ and\ \bibinfo {author} {\bibfnamefont {I.}~\bibnamefont {Riess}},\
  }\bibfield  {title} {\enquote {\bibinfo {title} {Application of percolation
  theory to 2d-3d heisenberg ferromagnets},}\ }\href@noop {} {\bibfield
  {journal} {\bibinfo  {journal} {Phys. Stat. Sol. (b)}\ }\textbf {\bibinfo
  {volume} {69}},\ \bibinfo {pages} {K15--K18} (\bibinfo {year}
  {1975})}\BibitemShut {NoStop}%
\bibitem [{\citenamefont {Granovetter}(1978)}]{Granovetter-1978}%
  \BibitemOpen
  \bibfield  {author} {\bibinfo {author} {\bibfnamefont {M.}~\bibnamefont
  {Granovetter}},\ }\bibfield  {title} {\enquote {\bibinfo {title} {Threshold
  models of collective behavior},}\ }\href@noop {} {\bibfield  {journal}
  {\bibinfo  {journal} {American J. Sociology}\ }\textbf {\bibinfo {volume}
  {83}},\ \bibinfo {pages} {1420--1443} (\bibinfo {year} {1978})}\BibitemShut
  {NoStop}%
\bibitem [{\citenamefont {Seidman}(1983)}]{Seidman-1983}%
  \BibitemOpen
  \bibfield  {author} {\bibinfo {author} {\bibfnamefont {S.~B.}\ \bibnamefont
  {Seidman}},\ }\bibfield  {title} {\enquote {\bibinfo {title} {Network
  structure and minimum degree},}\ }\href@noop {} {\bibfield  {journal}
  {\bibinfo  {journal} {Social Networks}\ }\textbf {\bibinfo {volume} {5}},\
  \bibinfo {pages} {269--287} (\bibinfo {year} {1983})}\BibitemShut {NoStop}%
\bibitem [{\citenamefont {Schwarz}\ \emph {et~al.}(2006)\citenamefont
  {Schwarz}, \citenamefont {Liu},\ and\ \citenamefont
  {Chayes}}]{Schwarz-Liu-Chayes-2006}%
  \BibitemOpen
  \bibfield  {author} {\bibinfo {author} {\bibfnamefont {J.~M.}\ \bibnamefont
  {Schwarz}}, \bibinfo {author} {\bibfnamefont {A.~J.}\ \bibnamefont {Liu}}, \
  and\ \bibinfo {author} {\bibfnamefont {L.~Q.}\ \bibnamefont {Chayes}},\
  }\bibfield  {title} {\enquote {\bibinfo {title} {The onset of jamming as the
  sudden emergence of an infinite $k$-core cluster},}\ }\href@noop {}
  {\bibfield  {journal} {\bibinfo  {journal} {Europhys. Lett.}\ }\textbf
  {\bibinfo {volume} {73}},\ \bibinfo {pages} {560--566} (\bibinfo {year}
  {2006})}\BibitemShut {NoStop}%
\bibitem [{\citenamefont {Sellitto}(2013)}]{Sellitto-2013}%
  \BibitemOpen
  \bibfield  {author} {\bibinfo {author} {\bibfnamefont {Mauro}\ \bibnamefont
  {Sellitto}},\ }\bibfield  {title} {\enquote {\bibinfo {title} {Disconnected
  glass-glass transitions and swallowtail bifurcations in microscopic spin
  models with facilitated dynamics},}\ }\href@noop {} {\bibfield  {journal}
  {\bibinfo  {journal} {J. Chem. Phys.}\ }\textbf {\bibinfo {volume} {138}},\
  \bibinfo {pages} {224507} (\bibinfo {year} {2013})}\BibitemShut {NoStop}%
\bibitem [{\citenamefont {Morone}\ \emph {et~al.}(2018)\citenamefont {Morone},
  \citenamefont {{Del Ferraro}},\ and\ \citenamefont
  {Makse}}]{Morone-etal-2018}%
  \BibitemOpen
  \bibfield  {author} {\bibinfo {author} {\bibfnamefont {F.}~\bibnamefont
  {Morone}}, \bibinfo {author} {\bibfnamefont {G.}~\bibnamefont {{Del
  Ferraro}}}, \ and\ \bibinfo {author} {\bibfnamefont {H.~A.}\ \bibnamefont
  {Makse}},\ }\bibfield  {title} {\enquote {\bibinfo {title} {The k-core as a
  predictor of structural collapse in mutualistic ecosystems},}\ }\href@noop {}
  {\bibfield  {journal} {\bibinfo  {journal} {Nature Phys.}\ }\textbf {\bibinfo
  {volume} {15}},\ \bibinfo {pages} {95--102} (\bibinfo {year}
  {2018})}\BibitemShut {NoStop}%
\bibitem [{\citenamefont {Kitsak}\ \emph {et~al.}(2010)\citenamefont {Kitsak},
  \citenamefont {Gallos}, \citenamefont {Havlin}, \citenamefont {Liljeros},
  \citenamefont {Muchnik}, \citenamefont {Stanley},\ and\ \citenamefont
  {Makse}}]{Kitsak-etal-2010}%
  \BibitemOpen
  \bibfield  {author} {\bibinfo {author} {\bibfnamefont {Maksim}\ \bibnamefont
  {Kitsak}}, \bibinfo {author} {\bibfnamefont {Lazaros~K.}\ \bibnamefont
  {Gallos}}, \bibinfo {author} {\bibfnamefont {Shlomo}\ \bibnamefont {Havlin}},
  \bibinfo {author} {\bibfnamefont {Fredrik}\ \bibnamefont {Liljeros}},
  \bibinfo {author} {\bibfnamefont {Lev}\ \bibnamefont {Muchnik}}, \bibinfo
  {author} {\bibfnamefont {H.~Eugene}\ \bibnamefont {Stanley}}, \ and\ \bibinfo
  {author} {\bibfnamefont {Hern\'{a}n~A.}\ \bibnamefont {Makse}},\ }\bibfield
  {title} {\enquote {\bibinfo {title} {Identification of influential spreaders
  in complex networks},}\ }\href@noop {} {\bibfield  {journal} {\bibinfo
  {journal} {Nature Phys.}\ }\textbf {\bibinfo {volume} {6}},\ \bibinfo {pages}
  {888--893} (\bibinfo {year} {2010})}\BibitemShut {NoStop}%
\bibitem [{\citenamefont {Chalupa}\ \emph {et~al.}(1979)\citenamefont
  {Chalupa}, \citenamefont {Leath},\ and\ \citenamefont
  {Reich}}]{Chalupa-Leath-Reich-1979}%
  \BibitemOpen
  \bibfield  {author} {\bibinfo {author} {\bibfnamefont {J.}~\bibnamefont
  {Chalupa}}, \bibinfo {author} {\bibfnamefont {P.~L.}\ \bibnamefont {Leath}},
  \ and\ \bibinfo {author} {\bibfnamefont {G.~R.}\ \bibnamefont {Reich}},\
  }\bibfield  {title} {\enquote {\bibinfo {title} {Bootstrap percolation on a
  bethe lattice},}\ }\href@noop {} {\bibfield  {journal} {\bibinfo  {journal}
  {J. Phys. C: Solid State Phys.}\ }\textbf {\bibinfo {volume} {12}},\ \bibinfo
  {pages} {L31--L35} (\bibinfo {year} {1979})}\BibitemShut {NoStop}%
\bibitem [{\citenamefont {Pittel}\ \emph {et~al.}(1996)\citenamefont {Pittel},
  \citenamefont {Spencer},\ and\ \citenamefont {Wormald}}]{Pittel-etal-1996}%
  \BibitemOpen
  \bibfield  {author} {\bibinfo {author} {\bibfnamefont {B.}~\bibnamefont
  {Pittel}}, \bibinfo {author} {\bibfnamefont {J.}~\bibnamefont {Spencer}}, \
  and\ \bibinfo {author} {\bibfnamefont {N.}~\bibnamefont {Wormald}},\
  }\bibfield  {title} {\enquote {\bibinfo {title} {Sudden emergence of a giant
  $k$-core in a random graph},}\ }\href@noop {} {\bibfield  {journal} {\bibinfo
   {journal} {J. Combin. Theory B}\ }\textbf {\bibinfo {volume} {67}},\
  \bibinfo {pages} {111--151} (\bibinfo {year} {1996})}\BibitemShut {NoStop}%
\bibitem [{\citenamefont {Dorogovtsev}\ \emph {et~al.}(2006)\citenamefont
  {Dorogovtsev}, \citenamefont {Goltsev},\ and\ \citenamefont
  {Mendes}}]{Dorogovtsev-etal-2006}%
  \BibitemOpen
  \bibfield  {author} {\bibinfo {author} {\bibfnamefont {S.~N.}\ \bibnamefont
  {Dorogovtsev}}, \bibinfo {author} {\bibfnamefont {A.~V.}\ \bibnamefont
  {Goltsev}}, \ and\ \bibinfo {author} {\bibfnamefont {J.~F.~F.}\ \bibnamefont
  {Mendes}},\ }\bibfield  {title} {\enquote {\bibinfo {title} {k-core
  organization of complex networks},}\ }\href@noop {} {\bibfield  {journal}
  {\bibinfo  {journal} {Phys. Rev. Lett.}\ }\textbf {\bibinfo {volume} {96}},\
  \bibinfo {pages} {040601} (\bibinfo {year} {2006})}\BibitemShut {NoStop}%
\bibitem [{\citenamefont {Baxter}\ \emph {et~al.}(2010)\citenamefont {Baxter},
  \citenamefont {Dorogovtsev}, \citenamefont {Goltsev},\ and\ \citenamefont
  {Mendes}}]{Baxter-etal-2010}%
  \BibitemOpen
  \bibfield  {author} {\bibinfo {author} {\bibfnamefont {G.~J.}\ \bibnamefont
  {Baxter}}, \bibinfo {author} {\bibfnamefont {S.~N.}\ \bibnamefont
  {Dorogovtsev}}, \bibinfo {author} {\bibfnamefont {A.~V.}\ \bibnamefont
  {Goltsev}}, \ and\ \bibinfo {author} {\bibfnamefont {J.~F.~F.}\ \bibnamefont
  {Mendes}},\ }\bibfield  {title} {\enquote {\bibinfo {title} {Bootstrap
  percolation on complex networks},}\ }\href@noop {} {\bibfield  {journal}
  {\bibinfo  {journal} {Phys. Rev. E}\ }\textbf {\bibinfo {volume} {82}},\
  \bibinfo {pages} {011103} (\bibinfo {year} {2010})}\BibitemShut {NoStop}%
\bibitem [{\citenamefont {Shrestha}\ and\ \citenamefont
  {Moore}(2014)}]{Shrestha-Moore-2014}%
  \BibitemOpen
  \bibfield  {author} {\bibinfo {author} {\bibfnamefont {M.}~\bibnamefont
  {Shrestha}}\ and\ \bibinfo {author} {\bibfnamefont {C.}~\bibnamefont
  {Moore}},\ }\bibfield  {title} {\enquote {\bibinfo {title} {Message-passing
  approach for threshold models of behavior in networks},}\ }\href@noop {}
  {\bibfield  {journal} {\bibinfo  {journal} {Phys. Rev. E}\ }\textbf {\bibinfo
  {volume} {89}},\ \bibinfo {pages} {022805} (\bibinfo {year}
  {2014})}\BibitemShut {NoStop}%
\bibitem [{\citenamefont {Rizzo}(2019)}]{Rizzo-2018}%
  \BibitemOpen
  \bibfield  {author} {\bibinfo {author} {\bibfnamefont {T.}~\bibnamefont
  {Rizzo}},\ }\bibfield  {title} {\enquote {\bibinfo {title} {Fate of the
  hybrid transition of bootstrap percolation in physical dimension},}\
  }\href@noop {} {\bibfield  {journal} {\bibinfo  {journal} {Phys. Rev. Lett.}\
  }\textbf {\bibinfo {volume} {122}},\ \bibinfo {pages} {108301} (\bibinfo
  {year} {2019})}\BibitemShut {NoStop}%
\bibitem [{\citenamefont {Altarelli}\ \emph
  {et~al.}(2013{\natexlab{a}})\citenamefont {Altarelli}, \citenamefont
  {Braunstein}, \citenamefont {{Dall'Asta}},\ and\ \citenamefont
  {Zecchina}}]{Altarelli-Braunstein-DallAsta-Zecchina-2013}%
  \BibitemOpen
  \bibfield  {author} {\bibinfo {author} {\bibfnamefont {F.}~\bibnamefont
  {Altarelli}}, \bibinfo {author} {\bibfnamefont {A.}~\bibnamefont
  {Braunstein}}, \bibinfo {author} {\bibfnamefont {L.}~\bibnamefont
  {{Dall'Asta}}}, \ and\ \bibinfo {author} {\bibfnamefont {R.}~\bibnamefont
  {Zecchina}},\ }\bibfield  {title} {\enquote {\bibinfo {title} {Optimizing
  spread dynamics on graphs by message passing},}\ }\href@noop {} {\bibfield
  {journal} {\bibinfo  {journal} {J. Stat. Mech.: Theor. Exp.}\ }\textbf
  {\bibinfo {volume} {2013}},\ \bibinfo {pages} {P09011} (\bibinfo {year}
  {2013}{\natexlab{a}})}\BibitemShut {NoStop}%
\bibitem [{\citenamefont {Altarelli}\ \emph
  {et~al.}(2013{\natexlab{b}})\citenamefont {Altarelli}, \citenamefont
  {Braunstein}, \citenamefont {{Dall'Asta}},\ and\ \citenamefont
  {Zecchina}}]{Altarelli-Braunstein-DallAsta-Zecchina-2013b}%
  \BibitemOpen
  \bibfield  {author} {\bibinfo {author} {\bibfnamefont {F.}~\bibnamefont
  {Altarelli}}, \bibinfo {author} {\bibfnamefont {A.}~\bibnamefont
  {Braunstein}}, \bibinfo {author} {\bibfnamefont {L.}~\bibnamefont
  {{Dall'Asta}}}, \ and\ \bibinfo {author} {\bibfnamefont {R.}~\bibnamefont
  {Zecchina}},\ }\bibfield  {title} {\enquote {\bibinfo {title} {Large
  deviations of cascade processes on graphs},}\ }\href@noop {} {\bibfield
  {journal} {\bibinfo  {journal} {Phys. Rev. E}\ }\textbf {\bibinfo {volume}
  {87}},\ \bibinfo {pages} {062115} (\bibinfo {year}
  {2013}{\natexlab{b}})}\BibitemShut {NoStop}%
\bibitem [{\citenamefont {Zhou}(2013)}]{Zhou-2013}%
  \BibitemOpen
  \bibfield  {author} {\bibinfo {author} {\bibfnamefont {H.-J.}\ \bibnamefont
  {Zhou}},\ }\bibfield  {title} {\enquote {\bibinfo {title} {Spin glass
  approach to the feedback vertex set problem},}\ }\href@noop {} {\bibfield
  {journal} {\bibinfo  {journal} {Eur. Phys. J. B}\ }\textbf {\bibinfo {volume}
  {86}},\ \bibinfo {pages} {455} (\bibinfo {year} {2013})}\BibitemShut
  {NoStop}%
\bibitem [{\citenamefont {Guggiola}\ and\ \citenamefont
  {Semerjian}(2015)}]{Guggiola-Semerjian-2015}%
  \BibitemOpen
  \bibfield  {author} {\bibinfo {author} {\bibfnamefont {A.}~\bibnamefont
  {Guggiola}}\ and\ \bibinfo {author} {\bibfnamefont {G.}~\bibnamefont
  {Semerjian}},\ }\bibfield  {title} {\enquote {\bibinfo {title} {Minimal
  contagious sets in random regular graphs},}\ }\href@noop {} {\bibfield
  {journal} {\bibinfo  {journal} {J. Stat. Phys.}\ }\textbf {\bibinfo {volume}
  {158}},\ \bibinfo {pages} {300--358} (\bibinfo {year} {2015})}\BibitemShut
  {NoStop}%
\bibitem [{\citenamefont {Yuan}\ \emph {et~al.}(2016)\citenamefont {Yuan},
  \citenamefont {Dai}, \citenamefont {Stanley},\ and\ \citenamefont
  {Havlin}}]{Yuan-etal-2016}%
  \BibitemOpen
  \bibfield  {author} {\bibinfo {author} {\bibfnamefont {X.}~\bibnamefont
  {Yuan}}, \bibinfo {author} {\bibfnamefont {Y.}~\bibnamefont {Dai}}, \bibinfo
  {author} {\bibfnamefont {H.~E.}\ \bibnamefont {Stanley}}, \ and\ \bibinfo
  {author} {\bibfnamefont {S.}~\bibnamefont {Havlin}},\ }\bibfield  {title}
  {\enquote {\bibinfo {title} {$k$-core percolation on complex networks:
  Comparing random, localized, and targeted attacks},}\ }\href@noop {}
  {\bibfield  {journal} {\bibinfo  {journal} {Phys. Rev. E}\ }\textbf {\bibinfo
  {volume} {93}},\ \bibinfo {pages} {062302} (\bibinfo {year}
  {2016})}\BibitemShut {NoStop}%
\bibitem [{\citenamefont {Pei}\ \emph {et~al.}(2017)\citenamefont {Pei},
  \citenamefont {Teng}, \citenamefont {Shaman}, \citenamefont {Morone},\ and\
  \citenamefont {Makse}}]{Pei-etal-2017}%
  \BibitemOpen
  \bibfield  {author} {\bibinfo {author} {\bibfnamefont {S.}~\bibnamefont
  {Pei}}, \bibinfo {author} {\bibfnamefont {X.}~\bibnamefont {Teng}}, \bibinfo
  {author} {\bibfnamefont {J.}~\bibnamefont {Shaman}}, \bibinfo {author}
  {\bibfnamefont {F.}~\bibnamefont {Morone}}, \ and\ \bibinfo {author}
  {\bibfnamefont {H.~A.}\ \bibnamefont {Makse}},\ }\bibfield  {title} {\enquote
  {\bibinfo {title} {Efficient collective influence maximization in cascading
  processes with first-order transitions},}\ }\href@noop {} {\bibfield
  {journal} {\bibinfo  {journal} {Sci. Rep.}\ }\textbf {\bibinfo {volume}
  {7}},\ \bibinfo {pages} {45240} (\bibinfo {year} {2017})}\BibitemShut
  {NoStop}%
\bibitem [{\citenamefont {Schmidt}\ \emph {et~al.}(2019)\citenamefont
  {Schmidt}, \citenamefont {Pfister},\ and\ \citenamefont
  {Zdeborov\'a}}]{Schmidt-etal-2018}%
  \BibitemOpen
  \bibfield  {author} {\bibinfo {author} {\bibfnamefont {C.}~\bibnamefont
  {Schmidt}}, \bibinfo {author} {\bibfnamefont {H.~D.}\ \bibnamefont
  {Pfister}}, \ and\ \bibinfo {author} {\bibfnamefont {L.}~\bibnamefont
  {Zdeborov\'a}},\ }\bibfield  {title} {\enquote {\bibinfo {title} {Minimal
  sets to destroy the $k$-core in random networks},}\ }\href@noop {} {\bibfield
   {journal} {\bibinfo  {journal} {Phys. Rev. E}\ }\textbf {\bibinfo {volume}
  {99}},\ \bibinfo {pages} {022310} (\bibinfo {year} {2019})}\BibitemShut
  {NoStop}%
\bibitem [{\citenamefont {Wang}\ \emph {et~al.}(2020)\citenamefont {Wang},
  \citenamefont {Cheng},\ and\ \citenamefont {Zhou}}]{Wang-etal-2020}%
  \BibitemOpen
  \bibfield  {author} {\bibinfo {author} {\bibfnamefont {S.-N.}\ \bibnamefont
  {Wang}}, \bibinfo {author} {\bibfnamefont {L.}~\bibnamefont {Cheng}}, \ and\
  \bibinfo {author} {\bibfnamefont {H.-J.}\ \bibnamefont {Zhou}},\ }\bibfield
  {title} {\enquote {\bibinfo {title} {Vulnerability and resilience of social
  engagement: Equilibrium theory},}\ }\href@noop {} {\bibfield  {journal}
  {\bibinfo  {journal} {Europhys. Lett.}\ }\textbf {\bibinfo {volume} {132}},\
  \bibinfo {pages} {60006} (\bibinfo {year} {2020})}\BibitemShut {NoStop}%
\bibitem [{\citenamefont {M\'ezard}\ \emph {et~al.}(2002)\citenamefont
  {M\'ezard}, \citenamefont {Parisi},\ and\ \citenamefont
  {Zecchina}}]{Mezard-etal-2002}%
  \BibitemOpen
  \bibfield  {author} {\bibinfo {author} {\bibfnamefont {M.}~\bibnamefont
  {M\'ezard}}, \bibinfo {author} {\bibfnamefont {G.}~\bibnamefont {Parisi}}, \
  and\ \bibinfo {author} {\bibfnamefont {R.}~\bibnamefont {Zecchina}},\
  }\bibfield  {title} {\enquote {\bibinfo {title} {Analytic and algorithmic
  solution of random satisfiability problems},}\ }\href@noop {} {\bibfield
  {journal} {\bibinfo  {journal} {Science}\ }\textbf {\bibinfo {volume}
  {297}},\ \bibinfo {pages} {812--815} (\bibinfo {year} {2002})}\BibitemShut
  {NoStop}%
\bibitem [{\citenamefont {Krzakala}\ \emph {et~al.}(2007)\citenamefont
  {Krzakala}, \citenamefont {Montanari}, \citenamefont {{Ricci-Tersenghi}},
  \citenamefont {Semerjian},\ and\ \citenamefont
  {Zdeborov\'a}}]{Krzakala-etal-PNAS-2007}%
  \BibitemOpen
  \bibfield  {author} {\bibinfo {author} {\bibfnamefont {F.}~\bibnamefont
  {Krzakala}}, \bibinfo {author} {\bibfnamefont {A.}~\bibnamefont {Montanari}},
  \bibinfo {author} {\bibfnamefont {F.}~\bibnamefont {{Ricci-Tersenghi}}},
  \bibinfo {author} {\bibfnamefont {G.}~\bibnamefont {Semerjian}}, \ and\
  \bibinfo {author} {\bibfnamefont {L.}~\bibnamefont {Zdeborov\'a}},\
  }\bibfield  {title} {\enquote {\bibinfo {title} {Gibbs states and the set of
  solutions of random constraint satisfaction problems},}\ }\href@noop {}
  {\bibfield  {journal} {\bibinfo  {journal} {Proc. Natl. Acad. Sci. USA}\
  }\textbf {\bibinfo {volume} {104}},\ \bibinfo {pages} {10318--10323}
  (\bibinfo {year} {2007})}\BibitemShut {NoStop}%
\bibitem [{\citenamefont {M{\'{e}}zard}\ and\ \citenamefont
  {Montanari}(2009)}]{Mezard-Montanari-2009}%
  \BibitemOpen
  \bibfield  {author} {\bibinfo {author} {\bibfnamefont {M.}~\bibnamefont
  {M{\'{e}}zard}}\ and\ \bibinfo {author} {\bibfnamefont {A.}~\bibnamefont
  {Montanari}},\ }\href@noop {} {\emph {\bibinfo {title} {Information, Physics,
  and Computation}}}\ (\bibinfo  {publisher} {Oxford Univ. Press},\ \bibinfo
  {address} {New York},\ \bibinfo {year} {2009})\BibitemShut {NoStop}%
\bibitem [{\citenamefont {Zhou}(2015)}]{Zhou-2015}%
  \BibitemOpen
  \bibfield  {author} {\bibinfo {author} {\bibfnamefont {H.-J.}\ \bibnamefont
  {Zhou}},\ }\href@noop {} {\emph {\bibinfo {title} {Spin Glass and Message
  Passing}}}\ (\bibinfo  {publisher} {Science Press},\ \bibinfo {address}
  {Beijing, China},\ \bibinfo {year} {2015})\BibitemShut {NoStop}%
\bibitem [{\citenamefont {Braunstein}\ \emph {et~al.}(2016)\citenamefont
  {Braunstein}, \citenamefont {Dall'Asta}, \citenamefont {Semerjian},\ and\
  \citenamefont {Zdeborov\'a}}]{Braunstein-etal-2016}%
  \BibitemOpen
  \bibfield  {author} {\bibinfo {author} {\bibfnamefont {A.}~\bibnamefont
  {Braunstein}}, \bibinfo {author} {\bibfnamefont {L.}~\bibnamefont
  {Dall'Asta}}, \bibinfo {author} {\bibfnamefont {G.}~\bibnamefont
  {Semerjian}}, \ and\ \bibinfo {author} {\bibfnamefont {L.}~\bibnamefont
  {Zdeborov\'a}},\ }\bibfield  {title} {\enquote {\bibinfo {title} {Network
  dismantling},}\ }\href@noop {} {\bibfield  {journal} {\bibinfo  {journal}
  {Proc. Natl. Acad. Sci. USA}\ }\textbf {\bibinfo {volume} {113}},\ \bibinfo
  {pages} {12368--12373} (\bibinfo {year} {2016})}\BibitemShut {NoStop}%
\bibitem [{\citenamefont {Zhou}(2016)}]{Zhou-2016b}%
  \BibitemOpen
  \bibfield  {author} {\bibinfo {author} {\bibfnamefont {H.-J.}\ \bibnamefont
  {Zhou}},\ }\bibfield  {title} {\enquote {\bibinfo {title} {A spin glass
  approach to the directed feedback vertex set problem},}\ }\href {\doibase
  10.1088/1742-5468/2016/073303} {\bibfield  {journal} {\bibinfo  {journal} {J.
  Stat. Mech.: Theor. Exp.}\ }\textbf {\bibinfo {volume} {2016}},\ \bibinfo
  {pages} {073303} (\bibinfo {year} {2016})}\BibitemShut {NoStop}%
\bibitem [{\citenamefont {Zhao}\ and\ \citenamefont
  {Zhou}(2016)}]{Zhao-Zhou-2016}%
  \BibitemOpen
  \bibfield  {author} {\bibinfo {author} {\bibfnamefont {J.-H.}\ \bibnamefont
  {Zhao}}\ and\ \bibinfo {author} {\bibfnamefont {H.-J.}\ \bibnamefont
  {Zhou}},\ }\href@noop {} {\enquote {\bibinfo {title} {Optimal discuption of
  directed complex networks},}\ }\bibinfo {howpublished} {eprint
  arXiv:1605.09257 [physics.soc-ph]} (\bibinfo {year} {2016})\BibitemShut
  {NoStop}%
\bibitem [{\citenamefont {Mugisha}\ and\ \citenamefont
  {Zhou}(2016)}]{Mugisha-Zhou-2016}%
  \BibitemOpen
  \bibfield  {author} {\bibinfo {author} {\bibfnamefont {S.}~\bibnamefont
  {Mugisha}}\ and\ \bibinfo {author} {\bibfnamefont {H.-J.}\ \bibnamefont
  {Zhou}},\ }\bibfield  {title} {\enquote {\bibinfo {title} {Identifying
  optimal targets of network attack by belief propagation},}\ }\href@noop {}
  {\bibfield  {journal} {\bibinfo  {journal} {Phys. Rev. E}\ }\textbf {\bibinfo
  {volume} {94}},\ \bibinfo {pages} {012305} (\bibinfo {year}
  {2016})}\BibitemShut {NoStop}%
\bibitem [{\citenamefont {Li}\ \emph {et~al.}(2021)\citenamefont {Li},
  \citenamefont {Zhang},\ and\ \citenamefont {Zhou}}]{Li-etal-2021}%
  \BibitemOpen
  \bibfield  {author} {\bibinfo {author} {\bibfnamefont {T.}~\bibnamefont
  {Li}}, \bibinfo {author} {\bibfnamefont {P.}~\bibnamefont {Zhang}}, \ and\
  \bibinfo {author} {\bibfnamefont {H.-J.}\ \bibnamefont {Zhou}},\ }\bibfield
  {title} {\enquote {\bibinfo {title} {Long-loop feedback vertex set and
  dismantling on bipartite factor graphs},}\ }\href@noop {} {\bibfield
  {journal} {\bibinfo  {journal} {Phys. Rev. E}\ }\textbf {\bibinfo {volume}
  {103}},\ \bibinfo {pages} {L061302} (\bibinfo {year} {2021})}\BibitemShut
  {NoStop}%
\bibitem [{\citenamefont {Liu}\ and\ \citenamefont
  {Barab\'asi}(2016)}]{Liu-Barabasi-2016}%
  \BibitemOpen
  \bibfield  {author} {\bibinfo {author} {\bibfnamefont {Y.-Y.}\ \bibnamefont
  {Liu}}\ and\ \bibinfo {author} {\bibfnamefont {A.-L.}\ \bibnamefont
  {Barab\'asi}},\ }\bibfield  {title} {\enquote {\bibinfo {title} {Control
  principles of complex systems},}\ }\href@noop {} {\bibfield  {journal}
  {\bibinfo  {journal} {Rev. Mod. Phys.}\ }\textbf {\bibinfo {volume} {88}},\
  \bibinfo {pages} {035006} (\bibinfo {year} {2016})}\BibitemShut {NoStop}%
\bibitem [{\citenamefont {Zdeborov\'a}\ \emph {et~al.}(2016)\citenamefont
  {Zdeborov\'a}, \citenamefont {Zhang},\ and\ \citenamefont
  {Zhou}}]{Zdeborova-etal-2016}%
  \BibitemOpen
  \bibfield  {author} {\bibinfo {author} {\bibfnamefont {L.}~\bibnamefont
  {Zdeborov\'a}}, \bibinfo {author} {\bibfnamefont {P.}~\bibnamefont {Zhang}},
  \ and\ \bibinfo {author} {\bibfnamefont {H.-J.}\ \bibnamefont {Zhou}},\
  }\bibfield  {title} {\enquote {\bibinfo {title} {Fast and simple decycling
  and dismantling of networks},}\ }\href@noop {} {\bibfield  {journal}
  {\bibinfo  {journal} {Sci. Rep.}\ }\textbf {\bibinfo {volume} {6}},\ \bibinfo
  {pages} {37954} (\bibinfo {year} {2016})}\BibitemShut {NoStop}%
\bibitem [{\citenamefont {Bau}\ \emph {et~al.}(2002)\citenamefont {Bau},
  \citenamefont {Wormald},\ and\ \citenamefont {Zhou}}]{Bau-Wormald-Zhou-2002}%
  \BibitemOpen
  \bibfield  {author} {\bibinfo {author} {\bibfnamefont {S.}~\bibnamefont
  {Bau}}, \bibinfo {author} {\bibfnamefont {N.~C.}\ \bibnamefont {Wormald}}, \
  and\ \bibinfo {author} {\bibfnamefont {S.}~\bibnamefont {Zhou}},\ }\bibfield
  {title} {\enquote {\bibinfo {title} {Decycling numbers of random regular
  graphs},}\ }\href@noop {} {\bibfield  {journal} {\bibinfo  {journal} {Random
  Struct. Alg.}\ }\textbf {\bibinfo {volume} {21}},\ \bibinfo {pages}
  {397--413} (\bibinfo {year} {2002})}\BibitemShut {NoStop}%
\bibitem [{\citenamefont {Zhao}(2017)}]{Zhao-2017}%
  \BibitemOpen
  \bibfield  {author} {\bibinfo {author} {\bibfnamefont {J.-H.}\ \bibnamefont
  {Zhao}},\ }\bibfield  {title} {\enquote {\bibinfo {title} {Generalized
  $k$-core pruning process on directed networks},}\ }\href@noop {} {\bibfield
  {journal} {\bibinfo  {journal} {J. Stat. Mech.: Theo. Exp.}\ }\textbf
  {\bibinfo {volume} {2017}},\ \bibinfo {pages} {063407} (\bibinfo {year}
  {2017})}\BibitemShut {NoStop}%
\bibitem [{\citenamefont {Xu}\ and\ \citenamefont {Zhou}(2017)}]{Xu-Zhou-2017}%
  \BibitemOpen
  \bibfield  {author} {\bibinfo {author} {\bibfnamefont {Y.-Z.}\ \bibnamefont
  {Xu}}\ and\ \bibinfo {author} {\bibfnamefont {H.-J.}\ \bibnamefont {Zhou}},\
  }\bibfield  {title} {\enquote {\bibinfo {title} {Optimal segmentation of
  directed graph and the minimum number of feedback arcs},}\ }\href@noop {}
  {\bibfield  {journal} {\bibinfo  {journal} {J. Stat. Phys.}\ }\textbf
  {\bibinfo {volume} {169}},\ \bibinfo {pages} {187--202} (\bibinfo {year}
  {2017})}\BibitemShut {NoStop}%
\end{thebibliography}

%

\end{document}